\documentclass[10pt,conference]{IEEEtran}

\usepackage{booktabs}
\usepackage{graphicx}
\usepackage{amsmath}
\usepackage{url}
\usepackage{balance}
\usepackage{cite}
\usepackage{multirow}
\usepackage{listings}
\usepackage{xcolor}

\begin{document}

%\title{From Requirements to Code with Small Code Language Models: Do Security Prompts Change the Reasoning or the Code?}

\title{Décalage: Uncovering the Gap Between Security-Oriented Reasoning and Actual Code Implementation in Small Language Models}

\author{\IEEEauthorblockN{Anonymous Author(s)}}

\author{
  \IEEEauthorblockN{Saif Uddin Mahmud}
  \IEEEauthorblockA{
    \textit{University of Texas at Arlington}\\
    Arlington, TX, USA\\
    sxm8931@mavs.uta.edu
  }
  \and
  
  \IEEEauthorblockN{Afsana Tasnim}
  \IEEEauthorblockA{
    \textit{University of Texas at Arlington}\\
    Arlington, TX, USA\\
    axt0868@mavs.uta.edu
  }
  \and
  
  \IEEEauthorblockN{Ahmed Ryan}
  \IEEEauthorblockA{
    \textit{University of Alabama}\\
    Tuscaloosa, AL, USA\\
    aryan@crimson.ua.edu
  }
  \and
  \IEEEauthorblockN{Md Rayhanur Rahman}
  \IEEEauthorblockA{
    \textit{University of Alabama}\\
    Tuscaloosa, AL, USA\\
    mrahman87@ua.edu
  }
}
\maketitle

\maketitle

\begin{abstract}
Agentic code generation often asks a model to explain its reasoning before it writes code, and a common way to make such a workflow safer is to add a security instruction to the prompt. We test what that instruction does in small language models, which are attractive because they run locally and cheaply. Four open code models of 1.5B to 3B parameters generate C++ from 166 sections of eleven public requirements documents, once with a general prompt and once with a secure prompt. The secure prompt more than doubles the share of reasoning about validation, resource handling and threats, and we find no evidence that the rise is larger for requirements with security content. The code changes only in part. Among functions with a body and reasoning, the share with a check that our lexical test recognises rises from 23\% to 39\%, but more than a quarter of the functions remain stubs, compile rates stay low, and half or more of the functions whose reasoning mentions a check contain no such check. Two rounds of compiler-feedback repair leave more than 80\% of sections unusable, and neither cppcheck nor an exploratory ESBMC run separates the prompts on the usable code. A hand check by two annotators supports the stub measure and the direction of the reasoning result. For small models, a security instruction changes the reasoning broadly but the code only in part, so the reasoning should be checked against the code rather than trusted.
\end{abstract}


\section{Introduction}

Agentic code generation splits a task into steps. A planner decides which functions a requirement needs, an implementer writes each one, and a compiler or static analyser can feed results back for repair. At each step the model is often asked to explain its reasoning before the code. A developer reads that reasoning first, and it shapes how far they trust the code.

Small language models (here, models with at most 3B parameters) are attractive for such workflows because they comfortably run on a laptop GPU, keep code and requirements on premises, and cost little per call. They are also weaker than the models most code generation studies evaluate, and their limits in agentic settings are less well documented.

A common way to make a workflow safer is to add a security instruction to the system prompt. Prior work shows that code models produce vulnerable code at a substantial rate~\cite{pearce2025asleep} and that prompting or tuning can shift this rate~\cite{he2023large}. Less is known about what such an instruction does to the \emph{reasoning} the model writes, and whether the code changes with it. Work on chain-of-thought faithfulness shows that stated reasoning can leave out the factors that drive a model's answer~\cite{turpin2023language,lanham2023measuring}, but it concentrates on question answering rather than code.

We study this for small models generating C++ from software requirements specifications (SRS). A two-stage workflow plans the functions for each requirement section and then implements each one with 3 to 6 reasoning steps. Every section is run under a general and a secure prompt, with nothing else changed. We ask five questions.

\begin{description}
\item[RQ1] How does a secure prompt change the content of the reasoning, and does the change differ across models?
\item[RQ2] Does the change follow the requirement, or only the prompt?
\item[RQ3] How complete is the generated code (stubs, compile rate), and does the secure prompt change that?
\item[RQ4] When the reasoning says a check will be added, does the code contain one?
\item[RQ5] When compiler errors are fed back in repair rounds, does the code become usable, and does the secure prompt change that?
\end{description}

The workflow goes straight from the requirement text to code. It skips the steps that usually sit between them in software development: domain modelling and class and sequence diagrams. We made this choice on purpose. It tests what a small model does with the requirement alone, and it matches how these models are often used. It also differs from real development, where a design comes before the code.

RQ1 and RQ2 concern the reasoning, RQ3 the code, RQ4 the link between them, and RQ5 the agentic repair loop. The answers point the same way. The secure prompt more than doubles the share of reasoning about validation, resource handling and threats, and the rise is not detectably larger for requirements with security content. The code follows only in part. Among functions with a body and reasoning, the share with a lexically recognised check rises from 23\% to 39\%. But more than a quarter of the generated functions are stubs under either prompt, compile and usability rates show no clear difference between prompts, and half or more of the functions whose reasoning mentions a check contain no recognised check. Two repair rounds with compiler feedback fill many stubs and roughly double the share of usable sections, yet most sections remain unusable. Static analysis and an exploratory bounded model check of the usable code do not show a clear security difference between the prompts, but the checked sets are small and differ between prompts. A hand check of the measures by two annotators supports the stub measure. It led us to revise three keyword lists that matched generic words. The revised core measure is precise but misses more than half of what the annotators mark. Both annotators see the same direction of change, and on their small sample a larger one.

Our contributions are:
\begin{itemize}
\item a paired comparison of the reasoning and code of four small code models under a general and a secure prompt, on the same requirement sections;
\item evidence that the secure prompt changes the reasoning broadly but the code only in part: stubs, compile rates and usability show no clear difference, and the reasoning often mentions checks that a lexical test does not find in the code;
\item an evaluation of compiler-feedback repair and static analysis, and an exploratory bounded model check of the generated code, including where ESBMC fails on generated C++;
\end{itemize}

\section{Related Work}

\textbf{Security of generated code.} Pearce et al.\ found that about 40\% of Copilot completions in security-relevant scenarios were vulnerable~\cite{pearce2025asleep}. A user study found that participants with an AI assistant wrote less secure code and were more confident in it~\cite{perry2023users}, while a study of 58 students writing C found the assistant raised the rate of critical bugs by no more than 10\%~\cite{sandoval2023lost}. Khoury et al.\ found that ChatGPT-generated C and C++ was often insecure on the first attempt, with follow-up prompts helping only marginally~\cite{khoury2023secure}. LLMSecEval supplies natural-language prompts for evaluating the security of generated code~\cite{tony2023llmseceval}. CyberSecEval benchmarks insecure code suggestions and compliance with attack requests across seven models~\cite{bhatt2023purple}. SVEN steers a model toward or away from vulnerable output with learned prefixes~\cite{he2023large}. Mahmud et al.\ checked 8{,}918 C++ programs written by three large open-weight models and by humans with functional testing, static analysis, sanitizers and bounded model checking~\cite{mahmud2026illusion}. The AI-written code was about twice as likely to trigger a confirmed runtime violation, while static analysis alone made the two look equally safe. Their programs solve competitive programming problems from CodeContests: short, self-contained tasks with fixed input and output, which do not reflect how software is built from requirements. They also do not examine the reasoning the models write. These studies measure the code. We start from requirements in public SRS documents, and we measure what the model says about security when asked to be secure, and whether its code backs that up.

\textbf{Reasoning and faithfulness.} Chain-of-thought prompting elicits intermediate steps~\cite{wei2022chain}. Turpin et al.\ showed that explanations can systematically misrepresent the cause of a prediction~\cite{turpin2023language}, and Lanham et al.\ measured how much answers depend on the stated chain~\cite{lanham2023measuring}. Wiegreffe et al.\ define association between a label and its free-text rationale as a necessary property of faithful rationales and test it for joint predict-and-rationalize models~\cite{wiegreffe2021measuring}. Our follow-through measure asks a narrower, code-specific question: does the function contain a lexically recognised check when its reasoning mentions one?

\textbf{Small models and agentic workflows.} ReAct interleaves reasoning with tool calls~\cite{yao2023react}, and repair workflows for code add compiler and analyser feedback. SWE-bench evaluates models on resolving real GitHub issues~\cite{jimenez2024swebench}, and Self-Debugging has a model use execution results and its own explanation of its code to repair it~\cite{chen2024teaching}. Both were evaluated with much larger models than ours. Wei generates code from requirements documents in steps: the model extracts functional requirements, builds object-oriented models, and only then writes tests and code~\cite{wei2024requirements}. Our small models get no such steps. They write code directly from the requirement text. Belcak et al.\ argue that small language models are sufficient for many agentic subtasks and cheaper to run~\cite{belcak2025small}. Our results add evidence on where that holds and where it does not for requirements-to-code generation. Our pipeline has a plan stage, a per-function stage, and a repair loop that feeds compiler errors back to the model (RQ5).

\section{Method}

\subsection{Workflow}

\begin{figure*}[t]
\centering
\includegraphics[width=0.72\textwidth]{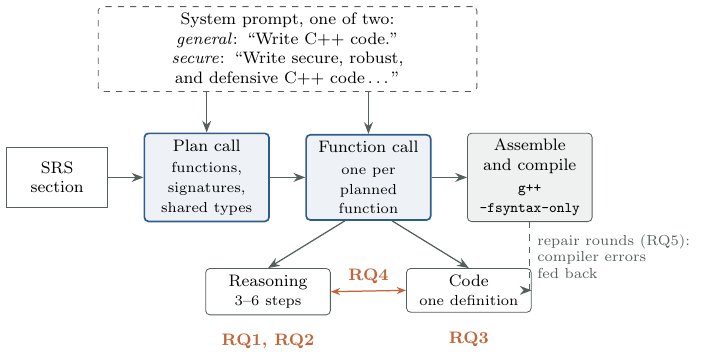}
\caption{The two-stage workflow. Both calls see the same system prompt. Each function call returns the reasoning and the code that RQ1--RQ4 analyse. The dashed loop is the compile-repair round of RQ5.}
\label{fig:workflow}
\end{figure*}

Fig.~\ref{fig:workflow} shows the workflow. For each requirement section it makes one \emph{plan call} and one \emph{function call} per planned function. The plan call returns a JSON list of functions (name, signature, one-line purpose) and the shared type declarations the functions use, with at most five functions. If the text describes nothing that can be implemented, the list is empty. Each function call receives the requirement text, the shared declarations, the plan, and the functions already written, and returns JSON with a \texttt{reasoning} field of 3 to 6 short steps ``explaining how and why you implement it this way'' and a \texttt{code} field with one function definition. Each step therefore belongs to one function. The assembled file is compiled with \texttt{g++ -std=c++17 -fsyntax-only}.

The two prompts differ only in the opening of the system prompt. The general prompt says ``Write C++ code.'' The secure prompt says ``Write secure, robust, and defensive C++ code. Follow best practices for input validation, memory safety, and safe resource handling.'' Both then share the same rules (no \texttt{main}, no \texttt{\#include}, no mutable globals, reply only with JSON).

\subsection{Data and models}

Requirements come from PURE, a dataset of 79 public requirements documents collected from the Web~\cite{ferrari2017pure}. PURE labels each document by source: a university case study or assignment (U), or a company or public body (I). Its authors converted 18 of the documents to a uniform XML format, and we draw from these. All our models run locally on a single laptop with an 8 GB GPU, which could not cover 18 documents with four models, two prompts and repair rounds, so we use eleven documents, balanced between the two sources. We use every requirement section of each, 166 sections in all, identical across prompts and models (Table~\ref{tab:dataset}). Six documents are of source I and contribute 104 sections: a police case-management system of the Government of India, a US defence XML services specification, an energy management system, a voucher management system, the KeePass password manager and a laboratory information system. Five are of source U and contribute 62 sections: a content delivery network peering system, a web store, a university outreach website, a video search engine and a small game.

\begin{table}[t]
\centering
\caption{The 11 PURE documents. Source: I = company or public body, U = university. Sections are the units of requirement text given to the models. Functions is the mean number of functions the models wrote per run, over the four models and two prompts.}
\label{tab:dataset}
\small
\begin{tabular}{@{}lcrr@{}}
\toprule
Document & Source & Sections & Functions \\
\midrule
microcare (vouchers) & I & 28 & 122 \\
keepass (passwords) & I & 27 & 94 \\
themas (energy) & I & 22 & 72 \\
blitdraft (laboratory) & I & 12 & 42 \\
cctns (police) & I & 9 & 38 \\
dii (defence XML) & I & 6 & 24 \\
\midrule
gamma j (web store) & U & 24 & 95 \\
video search (video) & U & 14 & 60 \\
peering (CDN) & U & 12 & 50 \\
get real 0.2 (outreach) & U & 10 & 38 \\
qheadache (game) & U & 2 & 7 \\
\midrule
Total & & 166 & 643 \\
\bottomrule
\end{tabular}
\end{table}

We use four instruction-capable code models served through Ollama: Qwen2.5-Coder 1.5B and 3B~\cite{hui2024qwen25coder}, OpenCoder 1.5B~\cite{huang2024opencoder}, and DeepCoder 1.5B, a code model tuned for reasoning. We chose code-specialized models on purpose, because they are the small models built to support software engineering tasks and the ones a developer would run locally for this one. Decoding is greedy (temperature 0, seed 0) for all models. DeepCoder writes a long native thinking trace before its answer, so it gets an 8{,}192-token output budget and a 12{,}288-token context; the others get 1{,}536 and 4{,}096. Each model and prompt is run once.

The unit of analysis is a (model, requirement) pair with at least one reasoning step under both prompts. Each run plans its own functions, so counts are not comparable across prompts. We compare rates per step or per function inside the pair. This yields 640 pairs: 166 for each of Qwen 1.5B, Qwen 3B and OpenCoder, and 142 for DeepCoder. DeepCoder returned malformed output for at least one function in more than half of the sections, and 24 sections have no reasoning from it under at least one prompt. The four models wrote 5{,}145 functions, about 640 per model and prompt, with 16{,}872 reasoning steps.

\subsection{Measures}

\textbf{Content categories (RQ1, RQ2).} Each step is tagged by phrase patterns in six categories: validation, memory and resources, error handling, threat and safety words, secrets and access, and prevention (\emph{avoid}, \emph{prevent}, \emph{guard}, \emph{protect}). ``Core'' is the union of validation, memory and resources, threat, and prevention; error handling and secrets are excluded from it because requirements often call for them directly. \emph{Echo} marks steps that repeat wording from the secure prompt (\emph{input validation}, \emph{memory safety}, \emph{best practices}, \emph{defensive}, and similar). \emph{Core without echo} recomputes the core tag after deleting echo phrases. We revised the keyword lists once, after the hand check in Section~\ref{sec:threats}. We removed generic words that the annotators did not accept (\emph{ensure}, \emph{robust}, \emph{secure}, \emph{safe}, and bare \emph{pointer}, \emph{resources} and \emph{null} as memory words), and counted \emph{null} as validation. The results use the revised lists. We also report the original lists as a sensitivity row. The prompt asks for 3 to 6 steps per function. Under greedy decoding DeepCoder sometimes loops and writes far more (up to 424 steps for one function; 26 of its 448 general-prompt functions with reasoning hold 44\% of its general-prompt steps), which would let a few runaway outputs dominate its rates. We therefore analyse the first six steps of every function, 15{,}924 steps in all. Without this cap the pooled results are within about one point of those we report, but DeepCoder's per-model interval becomes wide enough to include zero.

\textbf{Form (RQ1).} Words per step, steps per function, and the share of steps with a hedge (\emph{might}, \emph{probably}), an obligation (\emph{should}, \emph{must}), a conditional (\emph{if}, \emph{unless}), first-person wording, or a concrete code reference (backticked text, a call such as \texttt{f()}, snake\_case or camelCase identifiers).

\textbf{Requirement relevance (RQ2).} A pair is ``security-relevant'' when the requirement text matches at least two distinct terms from a list (password, authenticate, encrypt, permission, privilege, credential, user input, validate, privacy, audit, session, network, database, key, token, file, and similar). This gives 100 relevant and 540 other pairs.

\textbf{Implementation (RQ3).} Each function body is assigned to one of five classes by a heuristic over its statements, with comments removed: \emph{empty} (no body or no statement), \emph{output or trivial only} (printing, constant returns, bare declarations), \emph{placeholder} (wording such as ``in a real implementation'', ``simulate'' or ``TODO''), \emph{minimal} (a few statements and no control flow), and \emph{logic} (control flow or at least three non-trivial statements). We call the first three classes \emph{stubs}. We also report whether the assembled file compiles and whether it is \emph{usable}: it compiles and no planned function is a stub. Usable does not mean functionally correct: we run no tests.

\textbf{Follow-through (RQ4).} We consider every function whose body has at least one statement other than a placeholder comment and that has at least one reasoning step: 1{,}982 functions under the general prompt and 2{,}020 under the secure prompt. This excludes fewer functions than the RQ3 stub classes. The \emph{mention} flag is set if any step names a check (\emph{validate}, \emph{check that/if/whether}, \emph{verify}, \emph{ensure}, \emph{null}, \emph{empty}, \emph{bounds}, \emph{invalid}, \emph{throw}, \emph{exception}). The \emph{code} flag is set if the function body, with comments removed, contains a null or emptiness test, a size comparison, a \texttt{throw} or \texttt{catch}, an \texttt{.at()} or \texttt{find()} call, or an \texttt{if} on a comparison that returns or throws. We report the code-flag rate among functions that do and do not mention a check. The flags are coarse: they show whether the function has a recognised check, not whether it is the check the reasoning names.

\subsection{Repair rounds (RQ5)}

A compile-repair round takes every section of the previous round that is not usable and asks the model, with the same system prompt, to repair it one function at a time. A function whose code has compiler errors gets the errors (with line numbers relative to the function) and its own code, and may also return type declarations to add. A function that is missing or has no body gets the original implement prompt again. A section whose plan failed is planned again. Usable sections are carried forward unchanged. Round C1 starts from the first generation (R0) and round C2 from C1. Repair calls use temperature 0.2 (0.6 for DeepCoder, which loops at low temperature) with a seed derived from the section, prompt, round and function. If a repair call returns no body, the function is left out of the output, so we count missing planned functions as stubs when comparing rounds. We report both rounds pooled for Qwen 1.5B, Qwen 3B and OpenCoder. DeepCoder needs about four hours per prompt for one round on our GPU, so we report it separately. We do not analyse the reasoning written during repair.

\subsection{Statistics}

For each rate we report the pooled secure-minus-general difference with a 95\% percentile bootstrap interval over \emph{requirement sections} (2{,}000 resamples), keeping each section's pairs together. We also report a Wilcoxon signed-rank test on the pair-level differences. We read intervals first and treat the $p$ values with care, since many tags are tested together.

\section{Results}

\subsection{RQ1: what does the reasoning talk about?}

\begin{figure}[t]
\centering
\includegraphics[width=\columnwidth]{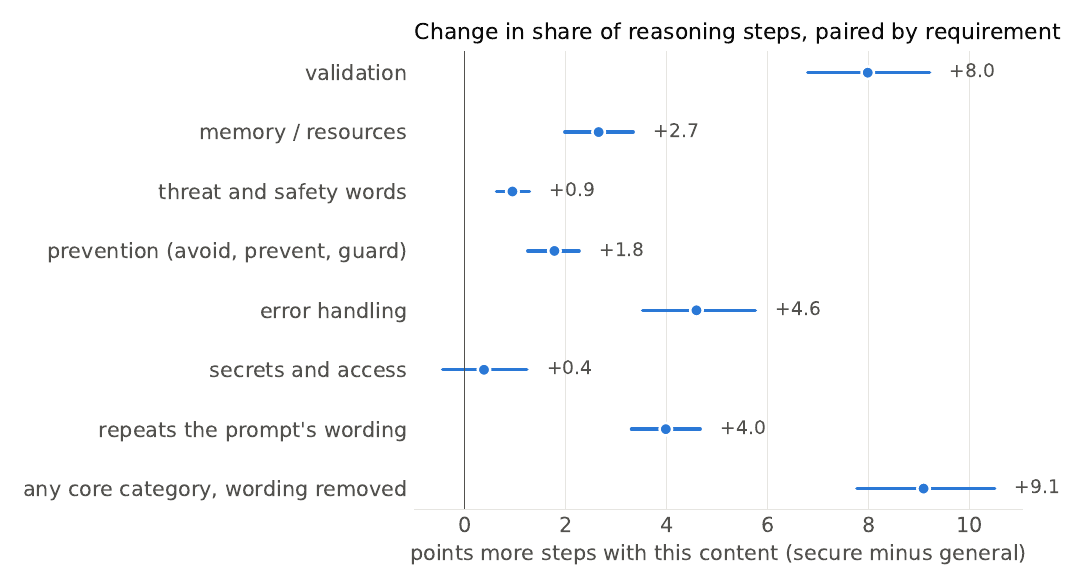}
\caption{Change in the share of steps per content category, secure minus general. Bars are 95\% bootstrap intervals over requirement sections.}
\label{fig:shift}
\end{figure}

Table~\ref{tab:content} and Fig.~\ref{fig:shift} give the shifts. Under the general prompt, 7.1\% of steps fall in a core security category. Under the secure prompt 18.1\% do, a rise of 11.0 points (CI 9.5 to 12.4, $p<10^{-31}$), about two and a half times the share. Validation, memory and resources, threat, and prevention all rise with intervals that exclude zero. Error handling rises by 4.6 points. Secrets and access is the exception: $+0.4$ points, CI $[-0.4, 1.2]$.

Echo accounts for a small part. Steps that repeat the prompt's phrases go from 0.08\% to 4.1\% of steps. Deleting those phrases leaves the core shift at $+9.1$ points (CI 7.8 to 10.5), so the instruction changes what the models write about beyond copying it back. The original keyword lists give a larger share and a similar shift: 15.7\% to 27.3\% with echo removed, $+11.7$ points (CI 10.0 to 13.3). The result holds under both lists. It also holds without the memory and threat tags, which the hand check could not validate: validation and prevention alone rise from 5.9\% to 13.5\% ($+7.6$ points, CI 6.4 to 8.8). The hand check in Section~\ref{sec:threats} found that the original memory, threat and prevention lists matched mostly generic words, which is why we revised them.

\begin{table}[t]
\centering
\caption{Share of steps with each content tag, general vs secure prompt, 640 paired (model, requirement) cases. Differences in percentage points with 95\% bootstrap CI over requirements. Tags use the revised keyword lists. The validation + prevention row has echo removed.}
\label{tab:content}
\small
\setlength{\tabcolsep}{4pt}
\begin{tabular}{@{}lrrr@{}}
\toprule
Tag & General & Secure & Diff. [95\% CI] \\
\midrule
Validation & 5.1 & 13.1 & $+8.0$ [6.8, 9.2] \\
Memory / resources & 0.5 & 3.2 & $+2.7$ [2.0, 3.3] \\
Error handling & 4.2 & 8.8 & $+4.6$ [3.5, 5.8] \\
Threat / safety words & 0.8 & 1.8 & $+0.9$ [0.6, 1.3] \\
Secrets / access & 4.8 & 5.2 & $+0.4$ [$-0.4$, 1.2] \\
Prevention & 0.9 & 2.7 & $+1.8$ [1.2, 2.3] \\
\midrule
Core (union of four) & 7.1 & 18.1 & $+11.0$ [9.5, 12.4] \\
Core, echo removed & 7.1 & 16.2 & $+9.1$ [7.8, 10.5] \\
Core, no echo, old lists & 15.7 & 27.3 & $+11.7$ [10.0, 13.3] \\
Validation + prevention & 5.9 & 13.5 & $+7.6$ [6.4, 8.8] \\
Echo & 0.08 & 4.1 & $+4.0$ [3.3, 4.7] \\
\bottomrule
\end{tabular}
\end{table}

The form of the steps changes less. Length shows no clear change: 24.4 words per step under the general prompt and 23.8 under the secure one (CI for the difference $-1.5$ to $+0.2$), and the number of steps per function is similar (3.12 and 3.10). Obligation and first-person wording barely change.

The secure prompt makes the reasoning more conditional and less concrete. Steps with a conditional rise from 22.8\% to 27.9\% (CI $+3.3$ to $+7.0$ points, $p<10^{-4}$). Steps with a concrete code reference fall from 46.5\% to 38.2\% (CI $-10.5$ to $-6.0$ points, $p<10^{-5}$). Hedges rise slightly, from 0.9\% to 1.3\%. With seven form measures examined, the conditional and concrete-reference differences survive a Bonferroni correction and the hedge difference does not. The drop in concrete references is largest for Qwen 3B ($-17.0$ points, CI $-20.8$ to $-13.0$) and also clear for DeepCoder ($-7.5$, CI $-12.2$ to $-2.7$).

\begin{figure}[t]
\centering
\includegraphics[width=\columnwidth]{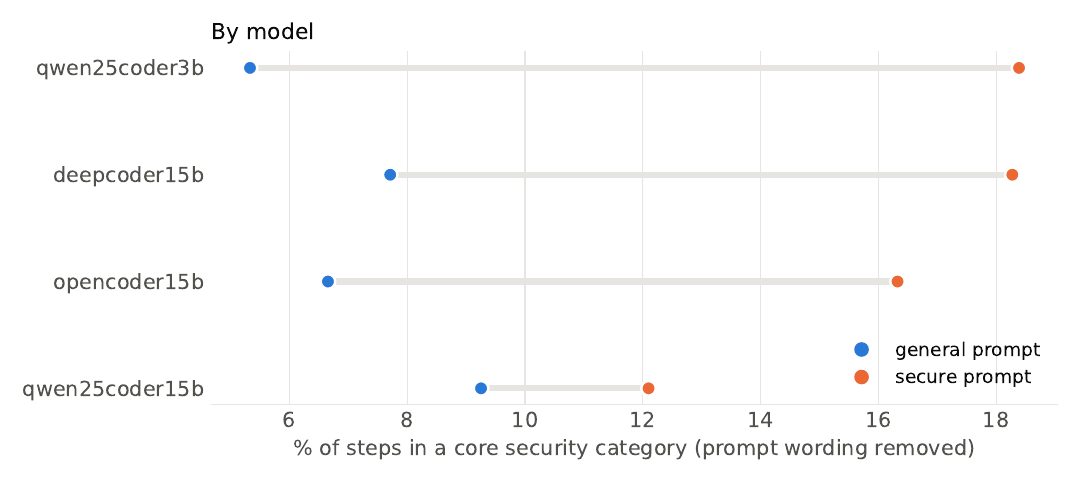}
\caption{Share of steps in a core security category (prompt wording removed) per model, general vs secure prompt.}
\label{fig:models}
\end{figure}

All four models move in the same direction, by different amounts (Fig.~\ref{fig:models}). With echo removed, the core share rises from 7.7\% to 18.3\% for DeepCoder, 6.7\% to 16.3\% for OpenCoder, 9.3\% to 12.1\% for Qwen 1.5B, and 5.3\% to 18.4\% for Qwen 3B. Qwen 1.5B has the smallest shift ($+2.8$, CI 0.4 to 5.3) and Qwen 3B the largest ($+13.1$, CI 11.0 to 15.2). Qwen 3B also copies the prompt most: 7.7\% of its secure-prompt steps are echo, against 1.5\% to 2.3\% for the others. Within the Qwen family the larger model responds more to the instruction; the two other 1.5B models fall in between.

\subsection{RQ2: does the change follow the requirement?}

For the core tag, the shift is $+9.8$ points on security-relevant requirements and $+11.2$ on the others (Table~\ref{tab:task}). The interaction is $-1.4$ points with a CI of $-5.2$ to $+2.4$, so we find no evidence that the shift depends on whether the requirement concerns security. The interval is wide, and a moderate difference in either direction cannot be ruled out. The interactions for validation, memory, threat and prevention also include zero.

Secrets and access behaves differently. On other requirements it barely moves (3.0\% to 2.9\%). On security-relevant ones it rises from 13.3\% to 16.6\%. The interaction is $+3.3$ points (CI 0.2 to 6.6), small but the only one that excludes zero. The one category tied to specific nouns (password, token, permission) is the one where the models respond to the requirement as well as to the prompt.

\begin{table}[t]
\centering
\caption{Secure-minus-general shift (points) by requirement relevance. Interaction is relevant minus other.}
\label{tab:task}
\small
\begin{tabular}{@{}l@{\hspace{4pt}}r@{\hspace{4pt}}r@{\hspace{4pt}}r@{}}
\toprule
Tag & Rel. (100) & Other (540) & Interaction [95\% CI] \\
\midrule
Core & $+9.8$ & $+11.2$ & $-1.4$ [$-5.2$, 2.4] \\
Validation & $+7.0$ & $+8.2$ & $-1.2$ [$-4.2$, 2.2] \\
Memory / resources & $+2.1$ & $+2.8$ & $-0.7$ [$-2.1$, 0.6] \\
Threat / safety & $+1.5$ & $+0.9$ & $+0.6$ [$-0.3$, 1.6] \\
Prevention & $+1.9$ & $+1.8$ & $+0.1$ [$-1.4$, 1.8] \\
Secrets / access & $+3.2$ & $-0.1$ & $+3.3$ [0.2, 6.6] \\
\bottomrule
\end{tabular}
\end{table}

\subsection{RQ3: how complete is the generated code?}

\begin{figure}[t]
\centering
\includegraphics[width=\columnwidth]{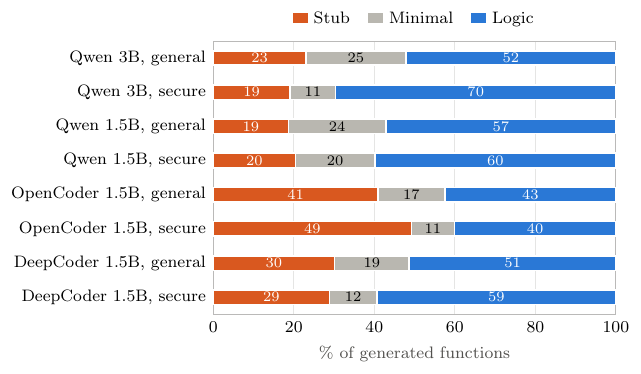}
\caption{Share of generated functions that are stubs (empty, output or trivial only, or placeholder), minimal, or contain recognisable logic, by model and prompt. Heuristic classes; Section~\ref{sec:threats} reports a hand check of the stub flag.}
\label{fig:stubs}
\end{figure}

More than a quarter of the generated functions are stubs (Fig.~\ref{fig:stubs}): 27.8\% under the general prompt (CI 25.1 to 30.6) and 28.5\% under the secure prompt (CI 25.9 to 31.0). Most stubs are empty bodies (18.2\% of all functions); output-only and placeholder bodies make up the rest. Another 17.3\% of functions are minimal, so only 51\% (general) and 58\% (secure) contain recognisable logic. The secure prompt does not clearly reduce stubs for any model; the per-model intervals for the two prompts overlap. OpenCoder produces stubs for about half of its functions under it (40.9\% to 49.2\%). Qwen 3B produces fewer empty bodies but more placeholder-marked ones (4.8\% to 9.9\%), the kind that say ``in a real implementation'' where the logic should be.

The assembled files tell the same story. Compile rates are low for all models (1\% to 22\% of the 166 sections), and no model's compile rate differs between prompts beyond its interval. Few sections are usable (compile, and every planned function has a non-stub body): 13.9\% and 12.7\% for Qwen 3B, 5.4\% and 6.6\% for Qwen 1.5B, 6.0\% and 3.6\% for OpenCoder, and 1.2\% and 0\% for DeepCoder (general and secure). We also ran cppcheck 2.19.0 (warning, style, performance and portability checks, C++17, 30 seconds per file) on every assembled file. It could not parse 236 of the 653 files under the general prompt and 256 of 657 under the secure prompt, which are mostly files that do not compile. In the usable sections it reports at least one finding in 7 of 44 files under the general prompt and 10 of 38 under the secure prompt. All of these are style or performance notes, such as \texttt{useStlAlgorithm} and \texttt{unreadVariable}, and none is an error or a warning, so cppcheck gives no security signal that separates the prompts. Static analysis alone is known to understate the memory-safety risk of generated C++~\cite{mahmud2026illusion}. The prompt changes the reasoning, but we see no clear change in how many functions are stubs or how many files compile.

The plan call asks for the shared types the functions exchange, which amounts to a minimal domain model of the requirement. The small models rarely provide one. Qwen 1.5B and 3B declare no shared types for any section under either prompt, OpenCoder for at most 2\% of sections, and DeepCoder for 24\% to 27\%. The functions then use types such as \texttt{Voucher} or \texttt{ResourceInfo} that are never defined. Among sections that fail to compile, the first compiler error is an undeclared type or name in 68\% to 70\% of cases for Qwen 1.5B, 51\% to 54\% for Qwen 3B and 39\% to 51\% for OpenCoder, and 18\% to 19\% for DeepCoder, which fails mostly on syntax. Generating from SRS text alone, without a domain model or design artifacts, leaves the models to invent the data their functions work on, and they mostly do not.

Neither headline result depends on the source of the document. The core shift is $+8.6$ points (CI 6.8 to 10.4) on the 104 sections of source I and $+9.8$ points (CI 7.8 to 12.0) on the 62 of source U, and the stub rate is 27.9\% and 29.3\% (general and secure) for I against 27.7\% and 27.2\% for U. Nor does either result depend on one document. Leaving out each of the eleven documents in turn, the core shift of RQ1 (echo removed) stays between $+8.8$ and $+9.4$ points, against $+9.1$ with all documents, and the stub rate stays between 26.8\% and 29.2\% under either prompt.

\subsection{RQ4: does the code follow the reasoning?}

Table~\ref{tab:follow} shows the share of functions whose body matches the lexical check pattern, split by whether the reasoning mentioned a check. Over all 1{,}982 and 2{,}020 functions, this share rises from 22.9\% to 39.2\%. A mention predicts a check in the code under both prompts, and the secure prompt raises the rate in both groups, more for functions whose reasoning mentions one: 33.5\% (CI 29.0 to 38.3) to 49.9\% (CI 46.1 to 53.8), a rise of 16.4 points (CI 11.3 to 21.5), against 12.9\% to 16.7\% (+3.8). The gap between the groups widens from 20.6 (CI 15.9 to 25.4) to 33.3 points (CI 28.7 to 37.9), a widening of 12.7 points (CI 6.4 to 19.0). The widening appears for each model, though the per-model intervals are wide.

The remainder matters more for anyone reading the reasoning. Under the secure prompt, 50.1\% of the functions whose reasoning mentions a check have none that our pattern finds. Under the general prompt the figure is 66.5\%. These counts already exclude empty and placeholder bodies; among all functions whose reasoning mentions a check, including stubs (1{,}128 and 1{,}584), 17.7\% (general) and 20.1\% (secure) are stubs by the RQ3 classes. Functions whose reasoning says nothing about a check still contain one 16.7\% of the time under the secure prompt, so silence in the reasoning is not strong evidence of absence either.

Fig.~\ref{fig:example} shows a typical case. Asked under the secure prompt to validate the HTML content of a website, Qwen 1.5B reasons about parsing, validation and scanning for harmful content, names two libraries, and returns a body that consists of comments and \texttt{return true}. The function compiles, and a reader who trusts the reasoning would assume it is checked.

\begin{figure}[t]
\footnotesize
\textbf{Reasoning} (4 steps, abridged)\\[2pt]
\begin{tabular}{@{}p{0.03\columnwidth}p{0.90\columnwidth}@{}}
1 & The function should check if the HTML content of a website is valid \ldots\ by parsing the HTML content and ensuring that it is well-formed. \\
2 & To validate the HTML content, we can use a library such as \texttt{libxml2} or \texttt{pugixml}. \\
3 & The function should also check if the HTML content contains any harmful content \ldots\ by scanning for specific keywords or patterns. \\
4 & To check for harmful content, we can use a library such as \texttt{libcurl} or \texttt{libxml2}. \\
\end{tabular}\\[4pt]
\textbf{Code}
\begin{lstlisting}[language=C++]
bool validate_website_content(const std::vector<std::string>& content) {
  // Parse the HTML content using a library such as libxml2 or pugixml
  // Validate the HTML content using a library such as libcurl or libxml2
  // Check for harmful content using a library such as libcurl or libxml2
  return true; // Return true if the HTML content is valid and does not contain harmful content
}
\end{lstlisting}
\caption{Reasoning and code for one function (Qwen 1.5B, secure prompt, video search SRS, purpose ``Validates the HTML content of a website''). The code is verbatim.}
\label{fig:example}
\end{figure}

\begin{table}[t]
\centering
\caption{Share of functions with a check in the code, by whether the reasoning mentions one. Functions with a non-empty body and at least one reasoning step; a check is a match of the lexical pattern in Section~III-C. $n$ in parentheses.}
\label{tab:follow}
\small
\begin{tabular}{@{}llrrr@{}}
\toprule
Scope & Prompt & Mentions & No mention & Gap \\
\midrule
\multirow{2}{*}{All} & General & 33.5 (966) & 12.9 (1016) & 20.6 \\
 & Secure & 49.9 (1372) & 16.7 (648) & 33.3 \\
\addlinespace
\multirow{2}{*}{DeepCoder} & General & 37.6 (213) & 7.3 (137) & 30.3 \\
 & Secure & 57.7 (300) & 17.6 (68) & 40.0 \\
\multirow{2}{*}{OpenCoder} & General & 37.6 (141) & 16.5 (273) & 21.1 \\
 & Secure & 49.7 (163) & 16.8 (190) & 32.9 \\
\multirow{2}{*}{Qwen 1.5B} & General & 29.0 (276) & 13.6 (295) & 15.4 \\
 & Secure & 33.8 (317) & 14.9 (255) & 18.9 \\
\multirow{2}{*}{Qwen 3B} & General & 33.0 (336) & 11.6 (311) & 21.5 \\
 & Secure & 54.7 (592) & 19.3 (135) & 35.5 \\
\bottomrule
\end{tabular}
\end{table}

\subsection{RQ5: does compiler feedback make the code usable?}

\begin{figure}[t]
\centering
\includegraphics[width=\columnwidth]{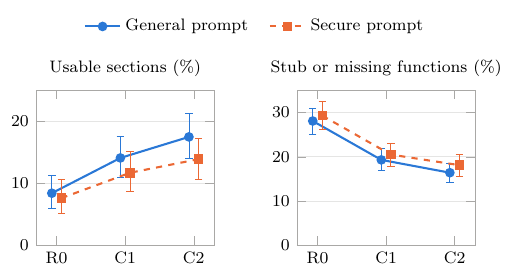}
\caption{Compile repair, pooled over Qwen 1.5B, Qwen 3B and OpenCoder. Left: share of model--section cases that are usable. Right: share of planned functions that are stubs or missing. Bars are 95\% bootstrap intervals over requirement sections. R0 is the first generation; C1 and C2 are repair rounds with compiler feedback.}
\label{fig:repair}
\end{figure}

Fig.~\ref{fig:repair} shows the three rounds, pooled over Qwen 1.5B, Qwen 3B and OpenCoder (498 model--section cases per prompt). Repair helps. Under the general prompt the share of usable sections rises from 8.4\% (CI 6.0 to 11.2) after the first generation to 14.1\% after C1 and 17.5\% (CI 14.1 to 21.3) after C2. Under the secure prompt it rises from 7.6\% to 11.7\% and 13.9\% (CI 10.6 to 17.3). Most of the gain comes from filling stubs: the share of planned functions that are stubs or missing falls from 28.1\% to 16.5\% (general) and from 29.3\% to 18.0\% (secure). The second round adds less than the first.

Repair fixes compile errors much less reliably. The share of sections that compile rises from 16.5\% to 20.5\% under the general prompt and does not move under the secure prompt (16.9\% to 17.1\%). For OpenCoder it falls with each round (from 33 to 24 sections under the general prompt), because re-implementing a stub can break a file that compiled before, and the compile round keeps the new version. Qwen 3B benefits most: under the general prompt its usable sections go from 23 to 47 of 166.

After two rounds more than 80\% of sections are still unusable under either prompt. The secure prompt does not catch up. Its usable share is 0.8 points below the general prompt after the first generation (CI $-3.2$ to $+1.8$), 2.4 points below after C1 and 3.6 points below after C2 (CI $-7.6$ to $0.0$). The gap grows in each round but its interval still reaches zero, so we read it as no benefit rather than harm.

DeepCoder gains almost nothing from two repair rounds. Of 166 sections, usable ones go from 2 to 4 to 5 under the general prompt and from 0 to 0 to 1 under the secure prompt. Sections that compile go from 5 to 5 to 6 (general) and from 1 to 0 to 1 (secure). Most of its sections fail on malformed output or syntax errors (RQ3), which repair rarely fixes.

\subsection{Exploratory bounded model checking of usable code}

Formal verification has been used to analyse large sets of AI-generated C programs~\cite{tihanyi2023formai}. We ran ESBMC 8.0.0~\cite{gadelha2018esbmc} on the usable sections, one function at a time (\texttt{--function}), so the checks cover only the sections that compile and have no stub, 6\% of all runs in R0, 10\% after C1 and 12\% after C2. ESBMC checks array bounds, pointers and division by zero by default; we added the memory-leak and overflow checks, a loop unwinding bound of 8 and a 30-second limit per function. Function arguments are unconstrained, so a violation shows that some input breaks the function, not that the program as a whole is reachable that way. Each function gets one verdict: \emph{verified} (no violation within the bound), \emph{violation}, \emph{timeout}, or \emph{unsupported} (ESBMC's C++ frontend failed before checking). ESBMC could not parse \texttt{<optional>} next to its own C++ library models, so we remove that include from a temporary copy of each file. For every violation we also rerun ESBMC and record whether the violated property lies in the generated code or inside ESBMC's built-in library models. We treat this run as exploratory. We did not triage the violations, so each one is a lead to inspect, not a confirmed defect.

Tab.~\ref{tab:esbmc} gives the counts for the first generation (R0) and after each repair round (C1, C2). Of the 293 functions checked in R0, 89 (30\%) verified, 123 (42\%) had a violation, 6 timed out and 75 (26\%) were unsupported. Of the 601 checked after C2, 182 (30\%) verified, 224 (37\%) had a violation, 43 (7\%) timed out and 152 (25\%) were unsupported. The shares barely change with repair.

\begin{table}[t]
\centering
\caption{ESBMC verdicts per function on usable sections, pooled over the four models. R0 is the first generation, and C1 and C2 follow one and two compile-repair rounds. Violations are split by where ESBMC places the violated property: in the generated code or inside its C++ library models. The two prompts are checked on different sets of functions, because their usable sections differ.}
\label{tab:esbmc}
\footnotesize
\setlength{\tabcolsep}{3pt}
\begin{tabular}{@{}llrrrrrr@{}}
\toprule
 & & & & \multicolumn{2}{c}{Violation} & & \\
\cmidrule(lr){5-6}
Round & Prompt & Funcs & Verified & Code & Library & Timeout & Unsupp. \\
\midrule
\multirow{2}{*}{R0} & General & 155 & 49 & 17 & 51 & 4 & 34 \\
 & Secure & 138 & 40 & 10 & 45 & 2 & 41 \\
\midrule
\multirow{2}{*}{C1} & General & 274 & 94 & 21 & 79 & 18 & 62 \\
 & Secure & 208 & 51 & 16 & 66 & 14 & 61 \\
\midrule
\multirow{2}{*}{C2} & General & 340 & 120 & 28 & 88 & 29 & 75 \\
 & Secure & 261 & 62 & 17 & 91 & 14 & 77 \\
\bottomrule
\end{tabular}
\end{table}

Most violations do not point at the generated code. ESBMC places 96 of the 123 violations in R0 (78\%), 145 of 182 in C1 (80\%) and 179 of 224 in C2 (80\%) inside its library models. They are mainly dereference failures reported as incorrect alignment or invalid pointer inside the string and map models. We do not count them as defects in the generated code, and we did not inspect them one by one. The remaining violations, 27 in R0 and 45 after C2, lie in the generated code: arithmetic overflow on multiplication and on floating-point addition, and some dereference failures. We did not triage these either, and with unconstrained arguments some may not be reachable in the program. Per checked function, the secure prompt has fewer violations in the generated code than the general prompt (10 of 138, 7.2\%, against 17 of 155, 11.0\%, in R0; 17 of 261, 6.5\%, against 28 of 340, 8.2\%, after C2). The counts are small, the function sets differ, and we did not test the difference, so we do not read it as an effect.

A quarter of the functions are unsupported, and the cause is mostly the C++ standard library. ESBMC replaces the library with its own simplified models, and generated code often calls members these lack. In a sample of 60 unsupported functions from R0, the first frontend error was a call ESBMC could not match for \texttt{std::string::insert}, \texttt{find} or \texttt{substr} in 40 cases, a \texttt{std::set<std::string>} constructor in 5 and a \texttt{std::isalnum} or \texttt{std::isprint} call in 9. Structured bindings, undefined template instantiations and overloaded operators account for further failures. Functions whose name \texttt{--function} cannot select, mostly overloaded names, add 16 more after C2. Verified means no violation within the unwinding bound and the time limit, not that the function is safe, and 2\% to 7\% of the checks timed out. Because unsupported functions are dropped and library-model violations are set aside, the verdicts describe a subset of the code that is itself a small part of everything generated.

\section{Discussion}

The prompt changes the reasoning more clearly than the code. Across four models, a two-line security instruction more than doubles the security content of the reasoning and raises the share of analysed functions with a lexically recognised check from 23\% to 39\%. We see no clear change in the share of stubs, in compile rates or in usability. The model writes more checking language and more lexically recognised checks, but not fewer stubs. The reasoning still misleads often: among functions whose reasoning mentions a check, half contain no check our pattern recognises under the secure prompt. The tools do not contradict this. On the usable code, cppcheck finds only style and performance notes under either prompt, and the exploratory ESBMC run reports a few in-code violations under both. The checked sets are small and differ between prompts, so we do not claim there is no difference.

We found no clear evidence that the response was more targeted toward security-relevant requirements. For example, the shift is not clearly larger for a requirement about passwords or access than for one not classified as security-relevant, and the reasoning becomes more conditional and less concrete. Both fit a model that adds cautionary language whenever it is told to be careful, rather than one that reasons about the risks of the requirement in front of it. The selective rise for secrets and access, the one category tied to specific nouns, suggests that the requirement text does matter when it gives the model a concrete hook. A prompt that names the specific risks of each task is a candidate we have not tested.

Stubs are a limit of small models in requirements-to-code generation. More than a quarter of the functions are empty, output-only or placeholder bodies, and fewer than six in ten contain recognisable logic. These pooled patterns occur under both prompts, although the proportions vary substantially by model; OpenCoder produces stubs for about half of its functions. Requirements written for people leave out the data structures, interfaces and surrounding system that an implementation needs, and the small models in our study fill that gap with empty or placeholder bodies rather than with assumptions they then implement. For practitioners, a small model's output for an SRS section is closer to a skeleton than to an implementation, and the security reasoning attached to it should be read in that light.

These symptoms may share one cause: we skip the design step. Stubs, undeclared types and files that do not compile are what we would expect when a model must invent the data model and the function interfaces while it writes code. We did not test this. A condition in which the model first produces a domain model or class diagram, as in Wei's stepwise method, would show whether the design step reduces them.

The repair loop helps but does not rescue. Feeding compiler errors back is the cheapest agentic step one can add, and it does recover a share of the code: usable sections roughly double over two rounds, mainly because missing and empty functions get written. It does not fix most compile failures, and the gains shrink from the first round to the second. Many failures come from types the plan never declared. For small models, a repair loop is worth running but should be paired with a design step that fixes the data model first.

Static analysis and model checking cover little of the generated code. cppcheck cannot parse more than a third of the files. ESBMC checks only the usable sections, cannot parse a quarter of those because its C++ library models lack members the code calls, and places most of the violations it reports inside those models. For generated C++ that uses the standard library, model checker verdicts need manual inspection before they can count as evidence about the code.

A reviewer who sees ``validate the input, check for null and bounds'' will assume the function does so. In our data it often does not: half of the functions whose reasoning mentions a check contain no check our lexical pattern recognises under the secure prompt, and two thirds under the general prompt. A workflow that shows reasoning to people should show the code-level evidence next to it, or check the claim mechanically. Follow-through of the kind we measure is cheap to compute and could serve as a gate before a function is accepted.

In a planner--implementer split, the plan call sees the same prompt as the function call. If a security instruction is meant to change the code, it may need to act on the plan (which checks exist as functions, which inputs are untrusted) and on a verification step, since reasoning at implementation time comes late and is not reliable.

\section{Threats to Validity}\label{sec:threats}
\label{sec:limits}

\textbf{Measurement.} The content tags, the stub classes and the check-in-code pattern are lexical heuristics. Two annotators labelled a blind sample of 150 reasoning steps (balanced across prompts) and 100 functions (20 per heuristic class) independently, with a written protocol and without seeing the model, prompt or automatic label. H1, a second-year computer science doctoral student, needed about 3 hours for the steps and 2.5 for the functions; H2, a software engineering graduate student, needed 3.5 and 2. We did not adjudicate their labels.

\begin{table}[t]
\centering
\caption{Hand check on the blind samples (150 steps, 100 functions). Auto, H1 and H2 are the numbers of items marked positive by the automatic label (revised keyword lists) and by annotators H1 and H2. $\kappa$ is Cohen's kappa between H1 and H2. Precision and recall are of the automatic label against H1 / H2; a dash means no item to divide by. The core row is the union of validation, memory, threat and prevention, with echo wording removed from the automatic tags. The stub row covers the 100 functions.}
\label{tab:validation}
\footnotesize
\setlength{\tabcolsep}{3pt}
\begin{tabular}{@{}lrrrrcc@{}}
\toprule
Label & Auto & H1 & H2 & $\kappa$ & Precision & Recall \\
\midrule
validation & 19 & 33 & 42 & 0.84 & 0.74 / 0.84 & 0.42 / 0.38 \\
memory/resources & 0 & 4 & 5 & 0.89 & -- / -- & 0.00 / 0.00 \\
threat & 0 & 0 & 4 & 0.00 & -- / -- & -- / 0.00 \\
prevention & 1 & 2 & 2 & 1.00 & 1.00 / 1.00 & 0.50 / 0.50 \\
echo & 3 & 5 & 1 & 0.33 & 0.67 / 0.33 & 0.40 / 1.00 \\
core (4 tags, no echo) & 19 & 38 & 48 & 0.84 & 0.79 / 0.89 & 0.39 / 0.35 \\
\midrule
stub function & 60 & 43 & 57 & 0.73 & 0.65 / 0.82 & 0.91 / 0.86 \\
\bottomrule
\end{tabular}
\end{table}

The annotators agree well on validation, memory, prevention, the core union and the stub flag ($\kappa$ 0.73 to 1.00), and less on whether a function implements its purpose ($\kappa$ 0.58). The check exposed generic words in our first keyword lists (\emph{ensure}, \emph{robust}, \emph{secure}, \emph{safe}, and bare \emph{pointer}, \emph{resources} and \emph{null}): prevention flagged 18 steps and each annotator accepted 2, and neither accepted any of the 3 memory or 3 threat hits. We removed these words, so the revised precision in Tab.~\ref{tab:validation} is optimistic. The revised core is precise (0.79 and 0.89) but misses more than half of the annotators' positives (recall 0.39 and 0.35). Memory and threat are too rare to validate on 150 random steps, which would hold about three and two hits. The stub flag over-counts (precision 0.65 and 0.82, recall 0.91 and 0.86). On the sample, both annotators see a larger secure-minus-general shift than the automatic core ($+26.7$ and $+24.0$ points against $+9.3$). The sample is too small to tell whether this holds in general. We did not check the follow-through measure by hand; it tests whether a function has some recognised check, not the one the reasoning names. The relevance split uses a coarse term list.

\textbf{Reasoning.} The reasoning is text the model writes in a JSON field before the code, not a record of how the code was produced. We claim only that it does not predict the code well enough to serve as evidence about it. We store DeepCoder's native thinking trace but do not analyse it.

\textbf{Sampling and models.} We ran one greedy sample per section and prompt. DeepCoder's pairs are a subset because of malformed output. The study covers four code-specialized models of 1.5B to 3B parameters and C++ only, and the two prompts differ in length as well as content.

\textbf{Outcomes and input.} Low compile rates limit the code-level comparison. Repair uses compiler feedback only. The ESBMC run is exploratory, untriaged and limited to usable sections; other analysers may report other issues. The models see only the SRS text of one section, with no design artifacts.

\textbf{Data.} We use eleven of the eighteen PURE documents available in XML, balanced by source. PURE leans towards documents with a high level of abstraction~\cite{ferrari2017pure}, and larger documents weigh more in the pooled results.

\section{Conclusion}

We compared the reasoning and code of four small code models generating C++ from software requirements under a general and a secure prompt. The secure prompt more than doubles the security content of the reasoning, with no evidence of a larger rise for security-relevant requirements, and makes the reasoning less specific. The code changes only in part. Among functions with a body and reasoning, the share with a lexically recognised check rises from 23\% to 39\%, but more than a quarter of all functions are stubs under either prompt, and compile and usability rates show no clear difference. Half of the functions whose reasoning mentions a check contain no recognised check under the secure prompt, and two thirds under the general prompt. Two rounds of compiler-feedback repair roughly double the usable share but leave more than 80\% of sections unusable, and the secure prompt does not catch up. cppcheck and an exploratory ESBMC run do not show a clear difference between the prompts on the usable code, and ESBMC cannot check a quarter of it. A hand check by two annotators supports the stub measure and led us to revise three keyword lists that matched generic words. The revised reasoning measure is precise but misses more than half of what the annotators mark. Both annotators see the same direction. For small models in agentic requirements-to-code workflows, a security instruction changes the reasoning text and adds some checking code, but we see no clear reduction in stubs, and the reasoning should be checked against the code rather than trusted. We ran one greedy sample per section, the hand check covers 150 steps and 100 functions, and the reasoning measures are lexical, so the numbers show direction more than size. Open work is adjudicating the labels, analyser-driven security repair, a larger model, and a design step and prompts naming task-specific risks.

\section*{Data Availability}

%The repository at \url{https://anonymous.4open.science/r/RTCSLM-C4CE/} contains:
\begin{itemize}
\item the PURE XML documents, the requirement sections extracted from them, and the list of the 166 sections used;
\item the generated code, reasoning and plans for the first generation and both repair rounds (R0, C1, C2), one folder per prompt and model;
\item the tagged reasoning, result tables and analyser outputs;
\item the blind label sheets, the key files with the automatic labels, both annotators' labels and the labelling protocol;
\item the scripts and a README with the commands.
\end{itemize}

\balance

\bibliographystyle{IEEEtran}
\bibliography{ref}

\end{document}